# Enhancement in transition temperature and critical current density of $CeO_{0.8}F_{0.2}FeAs$ by yttrium doping

J. Prakash[a], S.J. Singh[b], A. Banerjee[c], S. Patnaik[b]* and A.K. Ganguli[a]*,

[a]Department of Chemistry, Indian Institute of Technology, New Delhi 110016 India

[b]School of Physical Sciences, Jawaharlal Nehru University, New Delhi 110067 India

[c]UGC-DAE Consortium for Scientific Research, University Campus, Khandwa Road, Indore 452017 India

*Author for correspondence

Email: -   *spatnaik@mail.jnu.ac.in

　　　　  *ashok@chemistry.iitd.ernet.in

**Abstract**

We report significant enhancement in superconducting properties of yttrium substituted $Ce_{1-x}Y_xOFFeAs$ superconductors. The polycrystalline samples were prepared by two step solid state reaction technique. X-ray diffraction confirmed tetragonal ZrCuSiAs structure with decrease in both **a** and **c** lattice parameters on increasing yttrium substitution (with fixed F content). With smaller ion Y in place of Ce, the transition temperature increased by 6 K. Yttrium doping also lead to higher critical fields as well as stronger inter and intra-granular current density. The magnetization critical current density increased by an order of magnitude at 30 K and 1 T magnetic field.

Shortly after the discovery of superconductivity in oxypnictides [1], it was discerned that due to low transition temperature ($T_c$) and the deleterious influence of the grain boundaries, the possibility of technological application of these novel materials would be rather limited. The maximum $T_c$ ~55 K thus far is reported in a samarium - based compound [2] but this is far below the liquid nitrogen benchmark. Moreover the pnictides seem to have all the negatives of high $T_c$ cuprates particularly with respect to suppressed in-field intragranular pinning and weak link behavior across the grain boundaries [3-4]. One favorable aspect of oxypnictides is the high upper critical field ($H_{c2}$) as a consequence of multiband effects [5]. The important question of current relevance is the route to increase the transition temperature and critical current density in these materials without compromising on high upper critical field parameters. In this letter we report the successful synthesis of $Ce_{1-x}Y_xO_{0.8}F_{0.2}FeAs$ by simultaneous substitution of yttrium for cerium and fluorine for oxygen and study the superconducting properties in detail. We show that optimal Y addition can lead to significant increase in the transition temperature ($T_c$) and upper critical field ($H_{c2}$) as well as substantial enhancement of the critical current density ($J_c$) particularly at elevated temperatures.

Polycrystalline samples with nominal compositions of $Ce_{1-x}Y_xO_{0.8}F_{0.2}FeAs$ and $CeO_{0.8}F_{0.2}FeAs$ were synthesized by a two step solid state method [6] using high purity Ce, $CeO_2$, $Y_2O_3$, $CeF_3$ and FeAs as starting materials. FeAs was obtained by reacting Fe chips and As powder at 800 ºC for 24 hours. The raw materials were taken according to stoichiometric ratio and then sealed in evacuated silica ampoules ($10^{-4}$ torr) and heated at 900 °C for 30 hours. The powder was then compacted (5 tons) and the disks were wrapped in Ta foil, sealed in evacuated silica ampoules and heated at 1100 ºC for 30 h. All chemical manipulations were performed in a nitrogen-filled glove box. The samples were characterized by powder x-ray diffraction with Cu-

$K\alpha$ radiation. Resistivity measurements were carried out using a Cryogenic 8 T Cryogen-free magnet in conjunction with a variable temperature insert (VTI). The inductive part of the magnetic susceptibility was measured using a tunnel diode based RF (2.3 MHz) penetration depth technique [7]. Magnetization hysteresis loops and remanent magnetization of the samples were measured by a SQUID magnetometer.

Figure 1 shows the powder x-ray diffraction patterns for $Ce_{1-x}Y_xO_{0.8}F_{0.2}FeAs$ (x = 0.4) and $CeO_{0.8}F_{0.2}FeAs$. Majority of the observed reflections could be satisfactorily indexed on the basis of the tetragonal ZrCuSiAs type structure. Minor amount (~10%) of $Y_2O_3$ was observed as a secondary phase for Y-doped sample. The refined lattice parameters were found to be **a** = 3.9654(1) Å and **c** = 8.5803(3) Å for Y-doped sample and **a** = 3.988(3) Å and **c** = 8.607(8) Å for without Y-doped phase. The lattice parameters are smaller than the parent compound CeOFeAs (a = 3.996 Å and c = 8.648 Å [8]) and the reduction of the lattice volume upon F and Y-doping indicates a successful chemical substitution. Both the lattice parameters for $Ce_{1-x}Y_xOFFeAs$ (x = 0.1, 0.2, 0.3 and 0.4) decrease with the increase in yttrium content but in this work we focus only on x = 0 and x = 0.4 compositions. As shown in the inset of figure 1, the zero field onset transition temperature for x = 0.4 is estimated to be 48.6 K as compared to x = 0 case where it is 42.7 K [9]. Surprisingly for the Ce(O/F)FeAs, the $T_c$ is suppressed on increasing external pressure and is lowered to 1.1 K at 265 kbar [10]. On the contrary we find that the $T_c$ of Ce(O/F)FeAs enhances on increasing the chemical pressure by substituting smaller $Y^{3+}$ ion in place of $Ce^{3+}$. The onset of diamagnetic behavior below $T_c$ is confirmed from the inductive part of rf susceptibility as shown in the upper inset of figure 2. The onset transition temperature ($T_c$) determined is 47.5 K and 40.5 K for $Ce_{0.6}Y_{0.4}O_{0.8}F_{0.2}FeAs$ and $CeO_{0.8}F_{0.2}FeAs$ respectively. This bulk magnetization shows a slightly lower $T_c$ as compared to the resistivity data. The residual resistivity ratio (RRR = $\rho_{300} / \rho_{50}$) is estimated to be 6.6 and 3.5 for $CeO_{0.8}F_{0.2}FeAs$ and $Ce_{0.6}Y_{0.4}O_{0.8}F_{0.2}FeAs$ respectively. We emphasize that the onset of superconductivity in $Ce_{0.6}Y_{0.4}O_{0.8}F_{0.2}FeAs$ represents the highest transition temperature in Ce-based oxypnictides synthesized at ambient pressure.

The temperature dependence of the resistivity under varying magnetic fields is shown in figure 2 for sample $Ce_{0.6}Y_{0.4}O_{0.8}F_{0.2}FeAs$. Lower inset of figure 2 shows the zero field resistivity of the samples upto room temperature. From figure 2, it is clear that the $T_c$ (onset) shifts weakly with magnetic field, but the zero resistivity temperature shifts much more rapidly to lower temperatures. Using a criterion of 90 % of normal state resistivity, the estimated $H_{c2}$ -T phase diagram for both samples is shown in lower inset of figure 4. The WHH formula [11], gives $H_{c2}(0) = -0.693\ T_c\ (dH_{c2} / dT)_{T=Tc}$. The slope of $dH_{c2}/dT$ estimated from the $H_{c2}$-T plots is -2.67 and -1.45 T/K for Y-doped and without Y-doped samples respectively. Correspondingly, we find $H_{c2}(0) \sim 90$ T and 43 T for $Ce_{0.6}Y_{0.4}O_{0.8}F_{0.2}FeAs$ and $CeO_{0.8}F_{0.2}FeAs$ respectively. Evidently, we are able to significantly increase upper critical field by simultaneous doping of Y and F [12].

The inset in figure 3 shows a characteristic magnetization loop for the sample $Ce_{0.6}Y_{0.4}O_{0.8}F_{0.2}FeAs$. Both samples showed asymmetric M(H) loop similar to polycrystalline La iron oxypnictides [3]. The loop width ($\Delta m(H)$) is small, which could be due to either poor inter-grain connectivity or weak intragranular pinning [3]. The magnetic field dependence of the magnetization hysteresis loop widths ($\Delta m$) at 4.2 K, 15 K and 30 K for both samples are compared in figure 3. The samples were of irregular shape and we have normalized $\Delta m$ by the weight of the specimen. If approximated to a rectangular geometry, the Bean model critical current density $J_c$ (= $20\Delta m /V\ a\ (1-a/3b)$, where a is width, b is length and V is volume of sample) is estimated to be $3.9 \times 10^5$ and $1.2 \times 10^4$ A/m$^2$ for Ce$_{0.6}$Y$_{0.4}$O$_{0.8}$F$_{0.2}$FeAs and CeO$_{0.8}$F$_{0.2}$FeAs respectively at 30 K and 1T field. This needs comparison with both MgB$_2$ and cuprates [13-14] in a temperature and field range accessible by single stage cryo-coolers. Next we discuss whether this increase in J$_c$ with Y doping is due to increased intragranular pinning or better intergrain connectivity.

To analyze relative changes in inter and intra granular current density, in figure 4 we show the remanent moment as a function of increasing applied field. The samples were studied (at T = 5 K) under cycles of increasing magnetic field followed by removal of the field and measurement of remanent moment [3, 15]. When external field is applied, first the flux enters into the intergranular region followed by grain penetration above lower critical field H$_{c1}$. The remanent moment equals the product of local current density multiplied by current loop size and therefore a single step variation m$_R$ Vs. H curve indicates predominant role of intragrain supercurrents. We also observe that the remanent moment deviates from zero at ~ 7 mT and 20 mT for CeO$_{0.8}$F$_{0.2}$FeAs and Ce$_{0.6}$Y$_{0.4}$O$_{0.8}$F$_{0.2}$FeAs respectively. This implies some improvement in intergrain pinning as well in the Y doped sample. In the upper inset of figure 4 we plot H$_{c1}$(T) derived from the departure from linearity in low field M-H loop (elucidated in the inset of fig. 3). The technique is strictly applicable in slab geometry [16] and typically overestimates H$_{c1}$, but there is definite relative increase of the lower critical field for the Y doped sample.

In summary, we have demonstrated substantial enhancement in transition temperature, critical current density and critical fields by simultaneous doping of Y in place of Ce and F in place of O in semimetal CeOFeAs. This multitudinous beneficial effect is caused by three independent parameters; higher chemical pressure, selective tuning of interband scattering and superior pinning properties of Y$_2$O$_3$.

AKG and SP thank DST, Govt. of India for financial support. JP and SJS thank CSIR, Govt. of India, respectively for fellowships.

**Figure Captions**

**Figure 1:** Powder x-ray Diffraction patterns of (a) $Ce_{0.6}Y_{0.4}O_{0.8}F_{0.2}FeAs$ (b) $CeO_{0.8}F_{0.2}FeAs$. The impurity phases are $Fe_2As$ (+) and CeAs (&) for samples without Y and $Y_2O_3$ (*) for Y-doped sample. Inset of figure shows resistivity plot for $CeO_{0.8}F_{0.2}FeAs$ (○) and $Ce_{0.6}Y_{0.4}O_{0.8}F_{0.2}FeAs$ (●).

**Figure 2:** Temperature dependence of the electrical resistivity of $Ce_{0.6}Y_{0.4}O_{0.8}F_{0.2}FeAs$ under varying magnetic fields (0-4 T). Upper inset shows the inductive part of susceptibility as a function of temperature and lower inset represents the temperature dependence of resistivity up to room temperature for polycrystalline $CeO_{0.8}F_{0.2}FeAs$ (○) and $Ce_{0.6}Y_{0.4}O_{0.8}F_{0.2}FeAs$ (■).

**Figure 3:** Magnetic field dependence of magnetization hysteresis loop width (Δm) at 4.2, 15 and 30 K for $Ce_{0.6}Y_{0.4}O_{0.8}F_{0.2}FeAs$ (Closed symbols) in comparison with $CeO_{0.8}F_{0.2}FeAs$ (Open symbols). Inset shows the magnetization hysteresis loop at 4.2 K for polycrystalline $Ce_{0.6}Y_{0.4}O_{0.8}F_{0.2}FeAs$ and defines procedure for determination of $H_{c1}$.

**Figure 4:** The field-dependent remanent magnetization at 5 K for both samples. The upper inset shows the temperature dependence of lower critical field ($H_{c1}$) while the lower inset depicts the temperature dependence of upper critical field ($H_{c2}$) as a function of temperature for $Ce_{0.6}Y_{0.4}O_{0.8}F_{0.2}FeAs$ (■) and $CeO_{0.8}F_{0.2}FeAs$ (○).

**Figure1**.

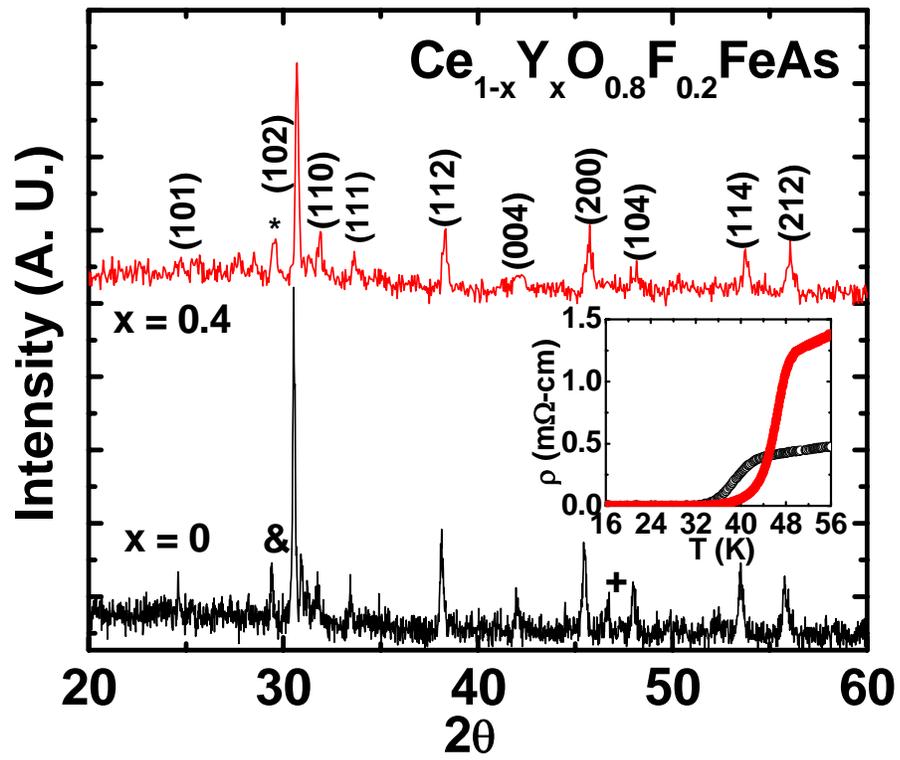

**Figure 2.**

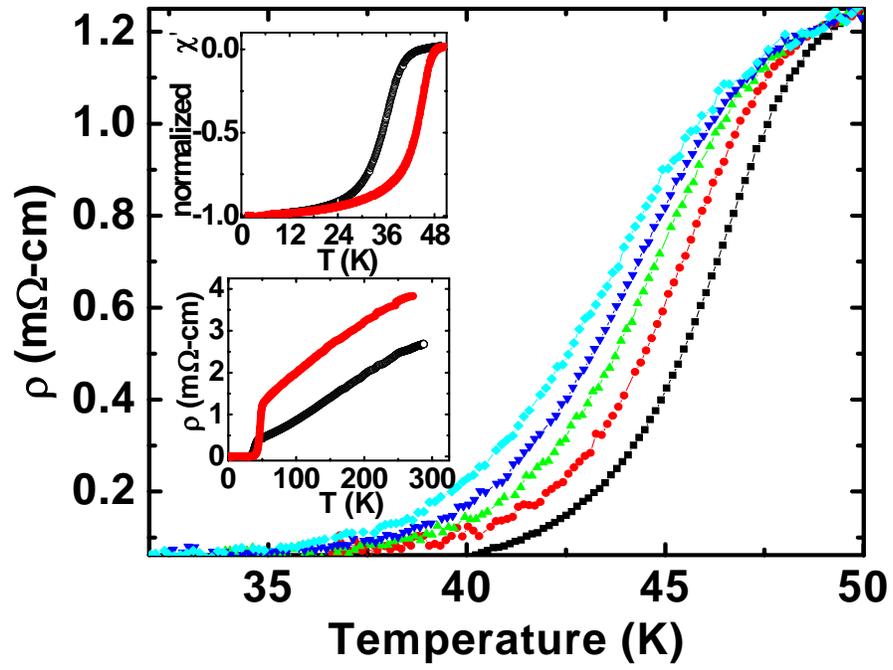

**Figure 3.**

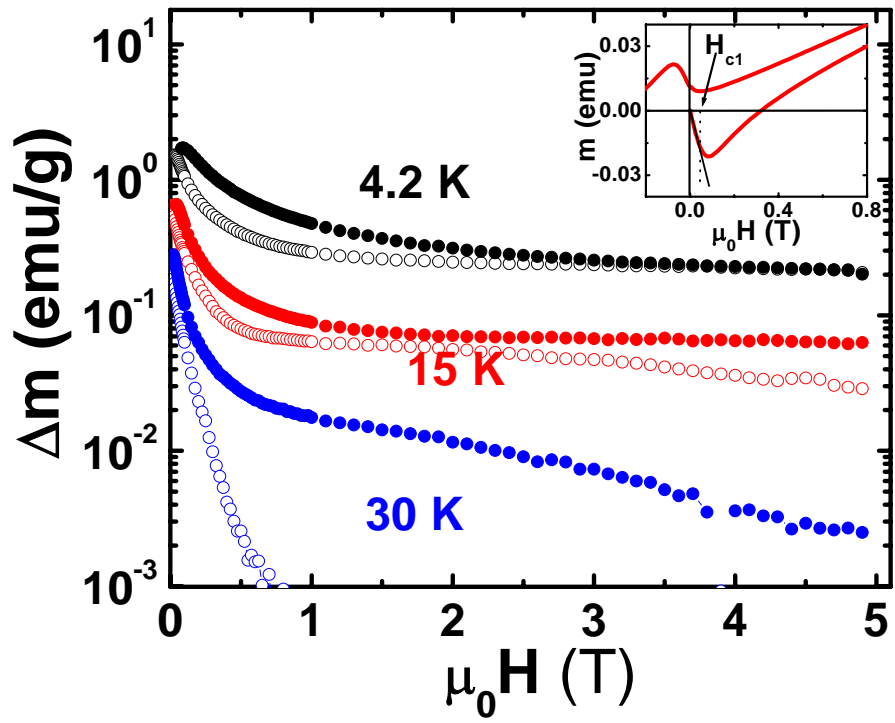

**Figure 4.**

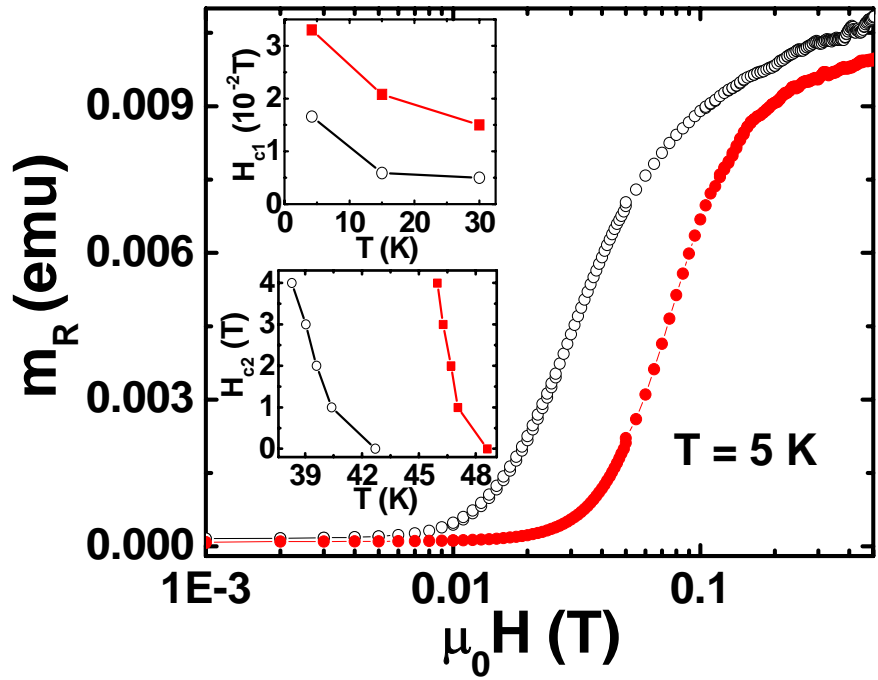